\documentclass[prb,twocolumn,showpacs,preprintnumbers,amsmath,amssymb]{revtex4}

\usepackage{graphicx}
\usepackage{dcolumn}
\usepackage{bm}

\newcommand{\mapright}[1]{\smash{\mathop{\hbox to 1.0cm{\rightarrowfill}}\limits^{#1}}}

\begin{document}

\preprint{}

\title{Effect of zero energy bound states on macroscopic quantum tunneling in high-$T_c$ superconductor junctions}

\author{Shiro Kawabata$^{1,2}$, Satoshi Kashiwaya$^3$, Yasuhiro Asano$^4$, and Yukio Tanaka$^5$}
\affiliation{%
$^1$Nanotechnology Research Institute and Synthetic Nano-Function Materials Project (SYNAF), National Institute of 
Advanced Industrial Science and Technology (AIST), Tsukuba, 
Ibaraki, 305-8568, Japan \\
$^2$International Project Center for Integrated Research on Quantum Information and Life Science (MILq Project), Hiroshima University, Higashi-Hiroshima, 739-8521, Japan\\
$^3$Nanoelectronics Research Institute, AIST, Tsukuba, 
Ibaraki, 305-8568, Japan \\
$^4$Department of Applied Physics, Hokkaido University,
Sapporo, 060-8628, Japan\\
$^5$Department of Applied Physics, Nagoya University,
Nagoya, 464-8603, Japan
}%

\date{\today}

\begin{abstract}
The macroscopic quantum tunneling (MQT) in the current biased high-$T_c$ superconductor Josephson junctions and the effect of the zero energy bound states (ZES) on the MQT are theoretically investigated.
We obtained the analytical formula of the MQT rate and showed that the presence of the ZES at the normal/superconductor interface leads to a strong Ohmic quasiparticle dissipation.
Therefore, the MQT rate is noticeably inhibited in compared with the $c$-axis junctions in which the ZES are completely absent.
\end{abstract}

\maketitle

Great attention has been attracted to theoretical and experimental studies of the effect of the dissipation on the macroscopic quantum tunneling (MQT) in superconductor Josephson junctions.~\cite{rf:MQT}
In current-biased Josephson junctions, the macroscopic variable is the phase difference between two superconductors and measurements of the MQT are performed by switching the junction from its metastable zero-voltage state to a non-zero voltage state (see Fig.~1 (c)).
Heretofore, experimental tests of the MQT have been focused on $s$-wave (low-$T_c$) junctions.~\cite{rf:DevoretReview,rf:sMQT1}
This fact is due to the naive preconception that the existence of the low energy nodal-quasiparticle in the $d$-wave order parameter of a high-$T_c$ cuprate superconductor~\cite{rf:d-wave1,rf:d-wave2} may preclude the possibility of observing the MQT.

Recently we have theoretically investigated the effect of the nodal-quasiparticle on the MQT in the $d$-wave $c$-axis junctions~\cite{rf:Kawabata1} (e.g., the Bi2212 intrinsic junction~\cite{rf:Intrinsic} and the cross-whisker junctions~\cite{rf:Takano}).
We have showed that the effect of the nodal-quasiparticle gives rise to a super-Ohmic dissipation~\cite{rf:d-waveaction1,rf:d-waveaction2} and the suppression of the MQT due to the nodal-quasiparticle is very weak. 
In fact, recently, Inomata $et$ $al.$ have experimentally observed the MQT in the Bi2212 intrinsic junctions.~\cite{rf:Inomata2}
They have reported that the effect of the nodal-quasiparticle on the MQT is negligibly small and the thermal-to-quantum crossover temperature is relatively high ($\sim1$K) in compared with the case of $s$-wave junctions.~\cite{rf:DevoretReview,rf:sMQT1}

In this paper, we will investigate the MQT in the $d$-wave junctions parallel to the $ab$-plane (see Fig.~1), e.g., the YBCO grain boundary junctions~\cite{rf:YBCOJJ1,rf:YBCOJJ2} and the ramp-edge junctions.~\cite{rf:YBCOJJ3}
In such junctions, the zero energy bound states (ZES)~\cite{rf:Hu,rf:TK95,rf:KashiwayaTanaka,rf:Lofwander} are formed near the interface between superconductor and the insulating barrier.
(Note that in the $d_0/d_0$ junction (Fig.~1 (a)) no ZES are formed as will be mentioned later.)
The ZES are generated by the combined effect of multiple Andreev reflections and the sign change of the $d$-wave order parameter symmetry, and are bound states for the quasiparticle at the Fermi energy.
Therefore the ZES may give rise to a strong dissipation for the MQT.
Note that recently Amin and Smirnov have theoretically calculated the decoherence time of a $d$-wave qubit and discussed the effect of the ZES on the qubit operation.~\cite{rf:Amin}
They, however, phenomenologically assumed that the system coupled to an Ohmic heat bath.
Instead, we will calculate the effective action starting from microscopic Hamiltonian without any phenomenological assumptions.
Moreover by using this effective action we will derive the theoretical formula of the MQT rate and discuss the influence of the ZES on the MQT.

%
%
%
%
\begin{figure}[b]
\begin{center}
\includegraphics[width=7.0cm]{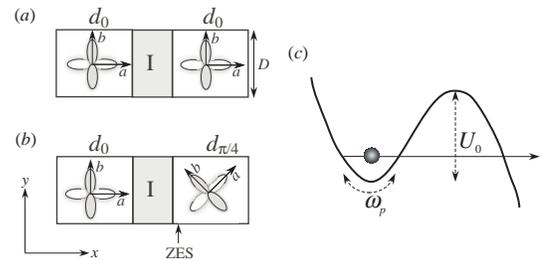}
\end{center}
\caption{Schematic drawing of the $d$-wave Josephson junction. (a) $d_0/d_0$ junction and (b) $d_0/d_{\pi/4}$ junction. 
(c) Potential $U(\phi)$ v.s. the phase difference  $\phi$ between two superconductors.
$U_0$ is the barrier height and $\omega_p$ is the Josephson plasma frequency.
}
\label{f1}
\end{figure}
%
%

The grandcanonical Hamiltonian of the $d$-wave junctions (Fig.~1 (a) and (b)) is given by
\begin{eqnarray}
{\cal H} =\sum_\sigma \int d \mbox{\boldmath $r$} 
\ 
\psi_{\sigma}^\dagger \left( \mbox{\boldmath $r$}  \right)
\left[
 - \frac{\hbar^2 \nabla^2 }{2 m} -\mu +W(\mbox{\boldmath $r$}) 
 \right]
\psi_{\sigma}\left( \mbox{\boldmath $r$}  \right)
 \nonumber\\
-
\frac{1}{2} \sum_{\sigma,\sigma'}  
\int d \mbox{\boldmath $r$}  d \mbox{\boldmath $r$}'
\ 
\psi_{\sigma}^\dagger \left( \mbox{\boldmath $r$}  \right)
\psi_{\sigma'}^\dagger \left( \mbox{\boldmath $r$}'  \right)
g \left( \mbox{\boldmath $r$}  - \mbox{\boldmath $r$}' \right)
\psi_{ \sigma'} \left( \mbox{\boldmath $r$}'  \right)
 \psi_{ \sigma} \left( \mbox{\boldmath $r$}  \right)
 ,
 \!  \! \!  \!  \!  \!  \!  \!  \!  
 \nonumber\\
 \end{eqnarray}
 where  $\mu$ is the chemical potential.
 This Hamiltonian describes conduction electrons in a potential $W(\mbox{\boldmath $r$})$ which includes a boundary.
 The second term in ${\cal H}$ describes the  anisotropic attractive interaction of strength $g(\mbox{\boldmath $r$})$. 
 This model also includes the insulating tunnel barrier I where $g(\mbox{\boldmath $r$})=0$ by a suitable choice of $W(\mbox{\boldmath $r$})$.
  Below,we will assume that the tunnel barrier is given by a delta function potential, i.e., $W(\mbox{\boldmath $r$})=w_0 \delta(x)$ and consider the high barrier limit $z_0 \equiv m w_0/\hbar^2 k_F \gg 1$ ($m$ is the mass of the electron and $k_F$ is the Fermi wave number.) which corresponds to typical experimental situations.

The partition function of the system can be written as a path integral over the Grasmmann fields,~\cite{rf:Functional} 
 \begin{eqnarray}
{\cal Z}
=
\int 
{\cal D} \bar{\psi} {\cal D} \psi 
\exp
\left(
  - \frac{1}{\hbar} \int_{0}^{\hbar \beta} d \tau {\cal L} [\tau]
\right)
,
\end{eqnarray}
where $\beta=1/k_BT$ and the Lagrangian is given by $ 
 {\cal L}[\tau]=\sum_\sigma\int d \mbox{\boldmath $r$} \bar{\psi}_\sigma (\mbox{\boldmath $r$},\tau) \partial_\tau  \psi_\sigma (\mbox{\boldmath $r$},\tau) + {\cal H}[\tau]
$.
In order to write the partition function as a functional integral over the macroscopic variable (the phase difference $\phi$), we apply the Hubbard-Stratonovich transformation.
This introduces a complex order parameter field $\Delta(\mbox{\boldmath $r$},\tau)$.
Now the integrals over the Grassmann fields  and $|\Delta|$ can be performed by using the Gaussian integral and the saddle point approximation, respectively.
Moreover, by following the method of Sch\"on and Zaikin,~\cite{rf:Schon1} 
 we obtain the partition function as $
 {\cal Z}
= 
\int 
{\cal D} \phi (\tau) 
\exp
\left(
  - {\cal S}_{eff}[\phi]/\hbar
\right)$,
where the effective action ${\cal S}_{eff}$  in the high barrier limit is given by 
\begin{eqnarray}
{\cal S}_{eff}[\phi]
= 
\int_{0}^{\hbar \beta} d \tau 
\left[
   \frac{M}{2} 
   \left(
   \frac{\partial \phi ( \tau) }{\partial \tau}
   \right)^2
   + 
   U(\phi)
\right]
\nonumber\\
-
\int_{0}^{\hbar \beta} d \tau 
\int_{0}^{\hbar \beta} d \tau'
\left[
  \alpha (\tau - \tau') \cos \frac{\phi(\tau) - \phi (\tau') }{2}
 \right]
.
\end{eqnarray}
In this equation, $M
 = 
 C \left( \hbar/2 e\right)^2
$
is the mass ($C$ is the junction capacitance) and the potential  $U(\phi)$ is described by
\begin{eqnarray}
 U(\phi) 
 = 
  \frac{\hbar}{2 e} 
\left[
    \int_0^1 d \lambda \ \phi  I_J (\lambda \phi) - \phi \  I_{ext}
\right]
,
\end{eqnarray}
where $I_J$ is the Josephson current and $ I_{ext}$ is the external bias current, respectively.
The dissipation kernel $\alpha(\tau)$ is related to the quasiparticle current $I_{qp}$
under constant bias voltage $V$ by
\begin{eqnarray}
\alpha(\tau) 
=
\frac{\hbar  }{e}
\int_0^\infty\frac{d \omega}{2 \pi}D_{\omega}(\tau) 
    I_{qp} \left( V=\frac{\hbar \omega}{e }\right) 
.
\end{eqnarray}
In this equation, $D_{\omega}(\tau)$ is the Matsubara Green's function of a free boson 
\begin{eqnarray}
D_{\omega}(\tau) 
=
\frac{1}{\hbar \beta} \sum_{n=-\infty}^\infty \frac{2 \hbar  \omega}{\nu_n^2 + (\hbar \omega)^2} \exp(i \nu_n \tau)
,
\end{eqnarray}
 where $\nu_n=2n \pi/\hbar \beta$ is the bosonic Matsubara frequency  ($n$ is an integer).

Below, we will derive the effective action for the two types of the $d$-wave junction ($d_0/d_0$ and $d_0/d_{\pi/4}$) in order to investigate the effect of the ZES on the MQT (see Fig.~1).
In the case of the $d_0/d_0$ junction, the node-to-node quasiparticle tunneling can contribute to the dissipative quasiparticle current.
However, the ZES are completely absent.
These behaviors are qualitatively identical with that for the $c$-axis Josephson junctions.~\cite{rf:Kawabata1}
On the other hand, in the case of the $d_0/d_{\pi/4}$ junction, the ZES are formed around the surface of the right superconductor $d_{\pi/4}$.
Therefore the node to the ZES quasiparticle tunneling becomes possible.

 Firstly, we will calculate the potential energy $U$ in the effective action (3).
 As mentioned above, $U$ can be described by the Josephson current through the junction.
 In order to obtain the Josephson current we start from the  Bogoliubov-de Gennes (B-dG) equation,~\cite{rf:KashiwayaTanaka} 

\begin{eqnarray}
&&
\! \! \! \! \! 
\int d \mbox{\boldmath $r$}'
\left(
\begin{array}{cc}
\delta(\mbox{\boldmath $r$} -\mbox{\boldmath $r$}' ) h(\mbox{\boldmath $r$}' )
& 
\Delta(\mbox{\boldmath $r$} -\mbox{\boldmath $r$}' )e^{i \varphi}
\\
\Delta^*(\mbox{\boldmath $r$} -\mbox{\boldmath $r$}' )e^{-i \varphi}
 &  
-\delta(\mbox{\boldmath $r$} -\mbox{\boldmath $r$}' ) h^*( \mbox{\boldmath $r$}' )
\end{array}
\right)
\left(
\begin{array}{c}
u({ \mbox{\boldmath $r$}}) \\
v({ \mbox{\boldmath $r$}})
\end{array}
\right)
\nonumber\\
&=&
E
\left(
\begin{array}{c}
u({ \mbox{\boldmath $r$}}) \\
v({ \mbox{\boldmath $r$}})
\end{array}
\right)
,
 \end{eqnarray}
where $\varphi$ is the phase of order parameter, $u(v)$ is the amplitude of the wave function for the electron (hole)-like quasiparticle, 
$
 h(\mbox{\boldmath $r$})
= - \hbar^2\nabla^2/2m -\mu+w_0 \delta(x)
$, and
$
\Delta(\mbox{\boldmath $r$} -\mbox{\boldmath $r$}' ) 
=\Omega^{-1} \sum_{\mbox{\boldmath $k$}} \Delta_{\mbox{\boldmath $k$}}
\exp \left[  i \mbox{\boldmath $k$} \cdot (\mbox{\boldmath $r$} -\mbox{\boldmath $r$}' ) \right]
$
($\Omega$ is the volume of the superconductor).
In the superconductor regions ($d_0$ and $d_{\pi/4}$), the B-dG equation (7) can be transformed into the eigenequation
\begin{eqnarray}
\left(
\begin{array}{cc}
\xi_{ \mbox{\boldmath $k$}} & \Delta_{ \mbox{\boldmath $k$}} e^{i \varphi}\\
 \Delta_{ \mbox{\boldmath $k$}} e^{-i \varphi}&  -\xi_{ \mbox{\boldmath $k$}}
\end{array}
\right)
\left(
\begin{array}{c}
u_{ \mbox{\boldmath $k$}} \\
v_{ \mbox{\boldmath $k$}}
\end{array}
\right)
=
E
\left(
\begin{array}{c}
u_{ \mbox{\boldmath $k$}} \\
v_{ \mbox{\boldmath $k$}}
\end{array}
\right)
,
 \end{eqnarray}
where, $\xi_{ \mbox{\boldmath $k$}}=\hbar^2 k^2/2m + \hbar^2 p^2/2m  - \mu \ $($p=2 \pi n/D$ and $D$ is the width of the junction).
The amplitude of the order parameter $\Delta_{\mbox{\boldmath $k$}}$ is given by $\Delta_0 \cos 2 \theta \equiv \Delta_{d_0}(\theta)$ for $d_0$ and $\Delta_0 \sin 2 \theta \equiv \Delta_{d_{\pi/4}}(\theta)$ for $d_{\pi/4}$, where $\cos \theta = k / k_F$.
The Andreev reflection coefficient for the electron (hole)-like quasiparticle $r_{he}$ ($r_{eh}$) is calculated by solving the eigenequation (8) together with the appropriate boundary conditions.
By substituting $r_{he}(r_{eh})$ into the formula of the Josephson current for unconventional superconductors (the Tanaka-Kashiwaya formula),~\cite{rf:KashiwayaTanaka,rf:TK96,rf:TK97,rf:Asano01} 
\begin{eqnarray}
I_J=
\frac{e  }{\hbar}
\sum_{p}
\frac{1}{\beta} \sum_{\omega_n}
\left(
\frac{\Delta_+}{\Omega_+} r_{he}
-
\frac{\Delta_-}{\Omega_-} r_{eh}
\right)
,
 \end{eqnarray}
we can obtain $\phi$ dependence of the Josephson current.
Here, $\Delta_\pm=\Delta_{(\pm k,p)}$, $\Omega_\pm = \sqrt{(\hbar \omega_n)^2 - |\Delta_\pm|^2}$, $\omega_n=(2n+1)\pi/\beta \hbar$ is the fermionic Matsubara frequency.
In the case of low temperatures ($\beta^{-1}\ll \Delta_0$), we get 
\begin{eqnarray}
 I_J(\phi) 
\approx 
\left\{
\begin{array}{cl}
\displaystyle{ I_1 \sin \phi}
 \quad 
 & \mbox{for} \quad  \mbox{$d_{0}/d_0$}
 \\
\displaystyle{- I_2 \sin2 \phi }
\quad
&
\mbox{for} \quad   \mbox{$d_{0}/d_{\pi/4}$}
\\
\end{array}
\right.
,
\end{eqnarray}
where $I_1 \equiv 3 \pi \Delta_0 / 10 e R_N$, $I_2 \equiv \pi^2 \hbar \beta \Delta_0^2 / 35 e^3 N_c R_N^2$ ($R_N=3 \pi \hbar z_0^2/ 2 e^2 N_c$ is the normal state resistance of the junction and $N_c $ is the number of channel at the Fermi energy).
Note that Eq.~(10) is only valid in the case of the high barrier limit ($z_0 \gg 1$).
By using Eq.~(4), we obtain the analytical expression of the potential $U,$ i.e.,
\begin{eqnarray}
U(\phi) 
\approx 
\left\{
\begin{array}{cl}
\displaystyle{- \frac{\hbar I_1}{2e}\left(  \cos \phi + \frac{I_{ext}}{I_1}  \phi \right) }
&    
\mbox{for} \ \   \mbox{$d_{0}/d_0$}
 \\
\displaystyle{- \frac{\hbar I_2}{4e}\left( -  \cos  2 \phi + 2 \frac{I_{ext}}{I_2} \phi  \right)}    
&    
\mbox{for} \ \   \mbox{$d_{0}/d_{\pi/4}$}
\end{array}
\right.
.
\end{eqnarray}
As in the case of the $s$-wave~\cite{rf:s-wave1,rf:s-wave2} and the $c$-axis junctions,~\cite{rf:Kawabata1} $U$ can be expressed as the tilted washboard potential (see Fig.~1(c)).

Next we will calculate the dissipation kernel $\alpha(\tau)$ in the effective action (3).
In the high barrier limit ($z_0 \gg 1$), the quasiparticle current is given by~\cite{rf:KashiwayaTanaka,rf:Lofwander}
\begin{eqnarray}
I_{qp}(V)
\! 
 &=& 
 \! 
 \frac{2e}{h} \sum_{p} |t_N|^2 \int_{-\infty}^{\infty}dE N_{L} (E,\theta)  N_{R} (E+ eV, \theta) 
\nonumber\\
&\times&
\left[
 f(E) - f(E +eV)
\right]
,
\end{eqnarray}
where $t_N \approx \cos \theta /z_0$ is the transmission coefficient of the barrier I, $N_{L(R)} (E,\theta)$ is the quasiparticle surface density of states ($L=d_0$ and $R=d_0$ or $d_{\pi/4}$), and $f(E)$ is the Fermi-Dirac distribution function.
The explicit expression of the surface density of states is obtained by Matsumoto and Shiba.~\cite{rf:Matsumoto}
In the case of $d_0$, there are no ZES.
Therefore the angle $\theta$ dependence of $N_{d_0}(E,\theta)$ is the same as the bulk and is given by
\begin{eqnarray}
 N_{d_{0}} (E,\theta)
=     \mbox{Re} \frac{|E|}{\sqrt{E^2-\Delta_{d_0}(\theta)^2}}
.
\end{eqnarray}
On the other hand, $N_{d_{\pi/4}}(E,\theta)$ is given by 
\begin{eqnarray}
\!
 N_{d_{\pi/4}} (E,\theta)
\!
=
\!
     \mbox{Re} \frac{\sqrt{E^2 \! - \! \Delta_{d_{\pi/4}}(\theta)^2}}{|E|} 
     \!
     +
     \!
      \pi |\Delta_{d_{\pi/4}}(\theta)| \delta(E)
.
\end{eqnarray}
The delta function peak at $E=0$ corresponds to the ZES. 
Because of the bound state at $E=0$, the quasiparticle current for the $d_0/d_{\pi/4}$ junctions is drastically different from that for the $d_0/d_0$ junctions in which no ZES are formed.
By substituting Eqs.~(13) and (14) into Eq.~(12), we can obtain the analytical expression of the quasiparticle current in the limit of low bias voltages ($e V \ll \Delta_0$) and low temperatures ($\beta^{-1} \ll \Delta_0$) as 
\begin{eqnarray}
I_{qp}(V) 
\approx 
\left\{
\begin{array}{cl}
\displaystyle{\frac{9 \pi^2}{256 \sqrt{2}} \frac{e V^2}{ \Delta_0 R_N } }
& 
\quad    \mbox{for} \quad \mbox{$d_{0}/d_{0}$}\\
\displaystyle{\frac{3 \pi^2}{16 \sqrt{2} } \frac{V}{  R_N } 
   }
& 
\quad   \mbox{for}\quad   \mbox{$d_{0}/d_{\pi/4}$}
\end{array}
\right.
.
\end{eqnarray}
Therefore from Eq.~(5), the asymptotic form ($\tau \gg \hbar/\Delta_0$) of the dissipation kernel for the $d$-wave junctions is given by 
\begin{eqnarray}
\alpha(\tau) 
\approx 
\left\{
\begin{array}{cl}
\displaystyle{\frac{9 }{128 \sqrt{2}}
\frac{\hbar^2 R_Q }{\Delta_0 R_N}
\frac{1}{|\tau|^3}
}
& 
\quad    \mbox{for} \quad \mbox{$d_{0}/d_{0}$}\\
\displaystyle{
\frac{3 }{16 \sqrt{2}}\frac{\hbar R_Q}{R_N}
\frac{1}{\tau^2}
}
&
\quad   \mbox{for}\quad   \mbox{$d_{0}/d_{\pi/4}$}
\end{array}
\right.
,
\end{eqnarray}
where $R_Q=h/4e^2$ is the resistance quantum.
These results (Eqs.~(15) and (16)) indicate that in the case of $d_0/d_0$ junctions, the dissipation is the super-Ohmic type as in the case of the $c$-axis junctions.~\cite{rf:Kawabata1}
This can be attributed to the effect of the node-to-node quasiparticle tunneling.
Thus the quasiparticle dissipation is very weak.
On the other hand, in the case of the $d_0/d_{\pi/4}$ junctions, the node-to-ZES quasiparticle tunneling gives the strong Ohmic dissipation which is similar to that in  normal junctions ($\alpha(\tau) \sim 1/\tau^2$).~\cite{rf:s-wave1,rf:s-wave2}
Therefore the dissipation for the $d_0/d_0$ junctions is enormously weaker than that for the $d_0/d_{\pi/4}$ junctions.

Let us move to the calculation of  the MQT rate $\Gamma$ for the $d$-wave Josephson junctions.
At the zero temperature $\Gamma$ is given by $\Gamma
=
\lim_{\beta \to \infty} (2/\beta) \ln{\cal Z} 
$.~\cite{rf:MQT}
Within the WKB approximation the partition function ${\cal Z}$ is evaluated by the saddle-point approximation.
Then $\Gamma$  in the low viscosity limit is obtained by a perturbative treatment~\cite{rf:Leggett} as $
\Gamma
\approx
A \exp \left(  - S_B/\hbar \right)$,
where $S_B \equiv {\cal S}_{eff}[\phi_B]$ and $\phi_B$ is the bounce solution.
Following the above procedures, we finally obtain the central results of this paper, i.e., the analytical formulae of the MQT rate for the $d$-wave junctions as~\cite{rf:Potential}
\begin{eqnarray}
\frac{\Gamma}{\Gamma_0}
&\approx&
    e^{
   -     \left[
     c \frac{81 \pi }{64 \sqrt{2}}  
       \frac{R_Q}{R_N}
       + 
       \frac{12}{5 }  \frac{\Delta_0 \delta M} {\hbar^2} 
        \right]
        \sqrt{ \frac{3 \pi }{5 } \frac{\hbar}{\Delta_0 C R_N}}
(1- x^2)^{\frac{5}{4}}   
}
    \nonumber\\
&& \quad    \mbox{for} \quad   d_{0}/d_0 ,
\\
\frac{\Gamma}{\Gamma_0}
&\approx&
     \exp 
    \left[
   -      \frac{81  \zeta (3)}{32 \sqrt{2} \pi^2 } \frac{R_Q}{R_N}
  (1- x^2)
    \right]
     \quad  \! \!  \! \!  \! \! \mbox{for} \! \! \! \!  \quad    d_{0}/d_{\pi/4}
     ,
\end{eqnarray}
where $c=\int_0^\infty d y \  y^4 \ln (1 + 1/y^2)/\sinh^2 (\pi y) \approx 0.0135$, $\zeta(3) \approx 1.20$ is the Riemann zeta function, $x=I_{ext}/I_{1(2)}$, and
$
\Gamma_0=12\omega_p 
\sqrt{
3U_0/ 2 \pi  \hbar \omega_p
  }
\exp 
\left( 
  -
36 U_0/ 5  \hbar \omega_p
 \right)
$
is the MQT rate without the dissipation ($U_0$ is the barrier height of the potential $U$ and $\omega_p$ is the Josephson plasma frequency. Note that $U_0$ and $\omega_p$ depend on the type of the junction).
In Eq.~(17) $\delta M$ is the renormalized mass due to the high frequency components ($\omega \geq \omega_p$) of the quasiparticle dissipation and is given by  
\begin{eqnarray}
\delta M 
&=&
\frac{3}{16\sqrt{2}} \frac{\hbar^2 R_Q}{\Delta_0 R_N}
\int_{-1}^1 d y  \ y^2 \frac{1+y}{\sqrt{1-y}}
\nonumber\\
&\times& 
\int_0^{\frac{\Delta_0}{\hbar \omega_p}} d z \ z^2 K_1 \left( z |y| \right)^2
,
\end{eqnarray}
where $K_1$ is the modified Bessel function.

In order to compare the influence of the ZES and the nodal-quasiparticle on the MQT more clearly, we will estimate the MQT rates (17) and (18) numerically.
For a mesoscopic bicrystal YBCO Josephson junction~\cite{rf:mesoYBCO} ($\Delta_0=17.8 \ $meV, $C=20 \times 10^{-15}\ $F, $R_N = 100  \ \Omega$A$x =0.95$), the MQT rate is estimated as 
\begin{eqnarray}
\frac{\Gamma}{\Gamma_0}
\approx
\left\{
\begin{array}{rl}
83 \% & \quad \mbox{for} \quad \mbox{$d_{0}/d_0$} \\
25 \% & \quad \mbox{for} \quad \mbox{$d_{0}/d_{\pi/4}$}
\end{array}
\right.
.
\end{eqnarray}
As expected, the node-to-ZES quasiparticle tunneling in the $d_{0}/d_{\pi/4}$ junctions gives strong suppression of the MQT rate in compared with the $d_{0}/d_{0}$ junction cases.

In conclusion, the MQT in the high-$T_c$ superconductors has been theoretically investigated.
The node-to-node quasiparticle tunneling in the $d_{0}/d_{0}$ junctions
gives rise to the weak super-Ohmic dissipation as in the case of the $c$-axis junctions.~\cite{rf:Kawabata1}
 For the $d_{0}/d_{\pi/4}$ junctions, on the other hand, we have found that the node-to-ZES quasiparticle tunneling leads to the strong Ohmic dissipation.
We have also analytically obtained the formulae of the MQT rate which can be used to
analyze experiments.

  In the context of an application to the $d$-wave phase qubit,~\cite{rf:Fominov,rf:Amin,rf:Zagoskin,rf:Kawabata2} it is desirable to use the $d_0/d_0$ or the $c$-axis Josephson junctions in order to avoid the strong Ohmic dissipation.
   However the effect of the ZES can be abated by several mechanisms (e.g., by applying magnetic field or by a disorder in the interface).~\cite{rf:KashiwayaTanaka,rf:Lofwander}
 Therefore it is interesting to investigate the MQT of the $d_{0}/d_{\pi/4}$ junction in such situations.

 Finally, we would like to comment about a recent experimental research.
Bauch $et$ $al.$ have succeeded to observe the MQT in a YBCO grain boundary bi-epitaxial Josephson junction.~\cite{rf:Bauch}
In their junction, however, the dissipation is mainly due to not intrinsic mechanisms (the ZES and the nodal-quasiparticles) which discussed in this paper but an $extrinsic$ impedance in parallel to the junction.
Therefore the development of the junction fabrication and the measurement techniques will enable us to  directly compare our theory with experimental results.

We  would like to thank S. Abe, P. Delsing, N. Hatakenaka, K. Inomata, T. Kato, and A. Tanaka for useful discussions.
This work was partly supported by NEDO under the Nanotechnology Program.


\begin{thebibliography}{99}
\bibitem{rf:Potential}
 In calculation we have approximated the potential $U(\phi)$ as the quadratic plus cubic type potential.~\cite{rf:Leggett}
Therefore the expressions (17) and (18) are only valid in the case where $x$ is close to unity.
\end{thebibliography}
\end{document}